# Wearable Slot Antenna at 2.45 GHz for Off-Body Radiation: Analysis of Efficiency, Frequency Shift, and Body Absorption

Marta Fernandez,[1,2] Hugo G. Espinosa ,[2]* David V. Thiel,[2] and Amaia Arrinda[1]

[1]*Department of Communications Engineering, University of the Basque Country (UPV/EHU), Bilbao, Spain*

[2]*School of Engineering, Griffith University, Brisbane, Australia*

The interaction of body-worn antennas with the human body causes a significant decrease in antenna efficiency and a shift in resonant frequency. A resonant slot in a small conductive box placed on the body has been shown to reduce these effects. The specific absorption rate is less than international health standards for most wearable antennas due to small transmitter power. This paper reports the linear relationship between power absorbed by biological tissues at different locations on the body and radiation efficiency based on numerical modeling ($r = 0.99$). While the $-10$ dB bandwidth of the antenna remained constant and equal to 12.5%, the maximum frequency shift occurred when the antenna was close to the elbow (6.61%) and on the thigh (5.86%). The smallest change was found on the torso (4.21%). Participants with body-mass index (BMI) between 17 and 29 kg/m$^2$ took part in experimental measurements, where the maximum frequency shift was 2.51%. Measurements showed better agreement with simulations on the upper arm. These experimental results demonstrate that the BMI for each individual had little effect on the performance of the antenna.

**Keywords: wearable antennas; power absorbed; antenna efficiency; resonant frequency shift; specific absorption rate**

## INTRODUCTION

Wearable sensors are used in athletic monitoring and human monitoring in normal living. Wireless transmission off the body is usually a line-of-sight radio link. A communications transceiver conveys relevant information off the body to a coach, a television channel, and/or a data-logging facility to score, analyze, and suggest improvements in human activities [Armstrong, 2007; Lee and Chung, 2009; Pantelopoulos and Bourbakis, 2010; Yang, 2014]. Wearable sensors and communication transmitters must be physically small, low powered, and conform to the human body for wearers' comfort. The small size of the antenna and close proximity to the human body greatly reduce the radiation range. For example, Varnoosfaderani et al. [2015b] reported a decrease in the off-body range at 2.45 GHz from 20 m in free space to 3 m for a $+10$ dBm transmitter located in an arm band positioned on the upper arm to a far field receiver with sensitivity of $-95$ dBm.

The location of the sensor is particularly important in acceleration measurements for biomechanical analyses [Nordsborg et al., 2014; Sabti and Thiel, 2014]. Clearly, the movement of each limb and above/below each joint is different, so off-body communications might use a central node with other sensors distributed around the body and connected wirelessly to the central node. This is exemplified in Figure 1 with a central node located on the torso.

Antennas placed on or close to lossy materials show a decrease in efficiency and a shift in the resonant frequency, even though the directivity is increased. These three effects are undesirable as the

Grant sponsors: Basque Government; grant number: IT-683-13; Erasmus Mundus PhD and Postdoctoral exchange program.

Conflict of interest: None.

*Correspondence to: Hugo G. Espinosa, School of Engineering, Griffith University, Nathan Campus, Brisbane, Qld 4111, Australia. E-mail: h.espinosa@griffith.edu.au

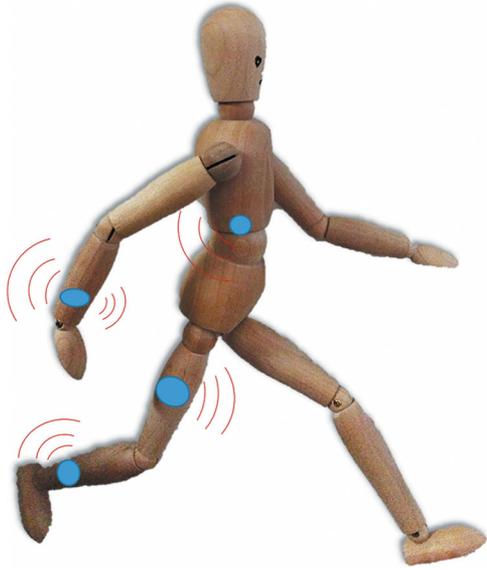

Fig. 1. Sensors connected from the body to an off-body receiver.

human body moves. For example, in Jiang and Werner [2015] the impact of the human body on input impedance of the antenna and radiation properties were studied when the antenna was located close to the arm and chest at a distance of approximately 2 mm. Two different body gestures were considered: standing and running modes. Frequency shifts and different peak gains could be observed due to body impact.

Any improvement in the antenna efficiency will increase transmission distance and/or battery life of wireless sensors. On many conductive surfaces, the antenna can be isolated from conducting materials using a ground plane. For wearable technology, this is not practical for reasons of human comfort and the separation distance required between the active element and ground plane. A flexible patch antenna is one strategy; however, the antenna center frequency and efficiency change with movement. A more efficient antenna design uses a rectangular resonant cavity with a slot [Takei et al., 1999; Varnoosfaderani et al., 2015a].

The power absorbed in human tissues is another key aspect of interest, as transmitters operate in close proximity to the body. Evaluation of the specific absorption rate (SAR) is crucial for compliance with safety levels assigned by international standards [ICNIRP, 1998; IEEE, 2005], and is often investigated [Anguera et al., 2012; De Santis et al., 2012; Risco et al., 2012; Soh et al., 2015]. Designing efficient antennas while maintaining a low SAR is one of the challenges in body area networks. Several antenna designs with low SAR have been proposed for 2.45 GHz band, which is suitable for on-body and off-body communications, for example, inverted-F antenna [Karmokar et al., 2010] and patch antenna [Rosaline and Raghavan, 2015]. In Soh et al. [2015], exposure of textile antennas was evaluated at different frequencies including 2.45 GHz, and they concluded that most measured SAR values were well below their respective simulated equivalent. A numerical solver was used to estimate worst-case on-body SAR.

This paper reports a redesigned cavity slot antenna operating at 2.45 GHz, and the relationship between total radiofrequency absorption and antenna efficiency when the antenna was mounted on different parts of the body. The three main features of this new antenna design are: the different feed probe, measurements on a variety of BMI humans, and the linear relationship between antenna efficiency and body absorption. The main advantage of this redesigned antenna is improvement of efficiency, which lies between 62% and 75% when it is placed on the body. In Varnoosfaderani et al. [2015a], the efficiency was 55% when the antenna was on the arm. The $S_{11}$ parameter was also improved compared to the previous design. Moreover, the slot dimensions of the new antenna ($47 \times 9$ mm) are significantly smaller than the previous design ($54 \times 30$ mm), and the material of the box in which the slot was printed is biodegradable polylactide (PLA). Specific absorption rate values were evaluated at 2.45 GHz in different simulation scenarios. The effect of positioning the antenna assembly on different locations on the body was investigated using participants with different BMIs. Practical information about experimental results and simulation accuracy compared with measurements is provided.

**THEORY AND MODELING**

A wire dipole antenna above a perfectly-conducting ground plane of infinite extent can be modeled as an image antenna [Balanis, 2008], and the radiation pattern can be calculated assuming two identical antennas in free space. As the distance above the ground plane decreases, the dipole impedance decreases until a zero separation distance, approaching the dipole antenna impedance to zero. If the ground plane is not perfectly conducting, it is possible to use complex image theory [Smith and King, 1981]. Given the complexity of the human anatomy and many body parts with different conductivity and permittivity, numerical modeling was implemented to investigate the performance of the antenna on the human body using commercially available finite-difference time-domain software (CST Computer Simulation Technology, Munich, Germany) [CST

Microwave Studio, 2016]. Antenna efficiency, various absorption coefficients, and radiation patterns of the antenna on human anatomy were determined.

The radiation efficiency $\eta$ in the vicinity of a series resonance of a generic antenna is defined by

$$\eta = \frac{R_r}{R_r + R_L} \quad (1)$$

where $R_r$ is the radiation resistance of the antenna and $R_L$ is the resistive loss in the antenna. The total absorbed power in biological tissue $P_a$ is given by

$$P_a = \int_V \sigma E^2 dV \quad (2)$$

where $\sigma$ is the conductivity of the tissue, and $E$ is the root mean square of the internal electric field generated within the tissue and contained in a volume element dV.

Radiation efficiency and power absorbed in the body are related by

$$\eta = \frac{P_r}{P_{in}} = \frac{P_r}{P_r + P_d + P_a} \quad (3)$$

where $P_r$ is radiated power, $P_{in}$ is input power, and $P_d$ is power dissipated in the antenna.

The SAR is a measure of the power absorbed per unit of mass and can be averaged over the whole body, or over a smaller part of the mass. In this study, SAR was averaged over 10 g of contiguous tissue and the maximum SAR reported for exposure at 2.45 GHz, as specified in ICNIRP [1998], IEEE [2005], and McIntosh and Anderson [2010], is given by

$$SAR = \frac{\sigma |E|^2}{\rho} \quad (4)$$

where $\rho$ is mass density of the biological tissue.

## MEASUREMENT TECHNIQUES

The effects of the human body on the performance of the antenna, when placed on different anatomical locations, can be analyzed using both the return loss and frequency shift. The return loss was obtained through simulation and experimental measurements. The dielectric properties of body tissues affect the performance of wearable antennas and influence power absorbed by the body. The dielectric properties of human tissues have been evaluated at different frequencies in past years [Gabriel et al., 1996; Gabriel and Gabriel, 1999]. However, the conductivity and permittivity values differ for every individual due to different factors such as anatomical aspects and age [Peyman et al., 2001; Vallejo et al., 2013]. At 2.45 GHz the conductivity and relative permittivity of the different body tissues vary from 0.095 to 3.458 S/m and from 5.147 to 68.361, respectively [IFAC, 1997]. For these reasons it is important to evaluate antenna properties not only on different participants but also on different parts of the body. People with different BMIs (17 and 29 kg/m$^2$) and different ages (22–59 years old) participated in the experiment. The locations on the body where the antenna was placed during simulations and measurements are indicated in Figure 2. These locations included several positions on the outer arm: on the wrist, above the elbow on the upper arm, on the middle of the torso close to the navel (Torso 1), on the left side of the torso (Torso 2), and on the thigh.

## ANTENNA DESIGN

A box made of biodegradable PLA material was fabricated using 3D printing technology (Bq Witbox 3D printer, Bq, Navarra, Spain) ($\varepsilon_r = 4$, $\tan\delta = 0.02$, where $\varepsilon_r$ is relative permittivity and $\tan\delta$ is dielectric loss tangent). The internal walls of the box were coated with conducting silver paste (Acheson

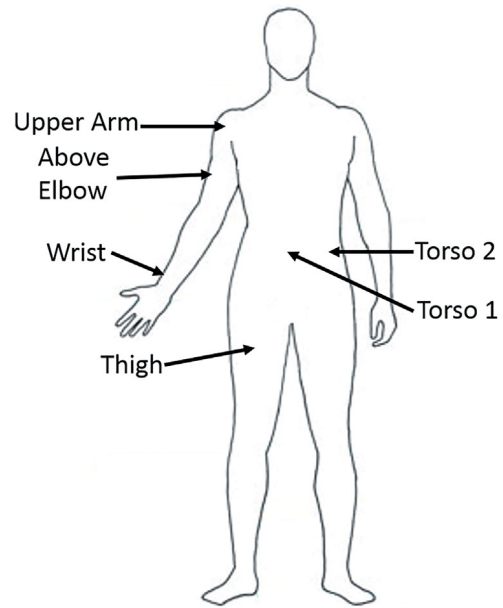

Fig. 2. Positions of the body where antenna was placed in simulations and measurements.

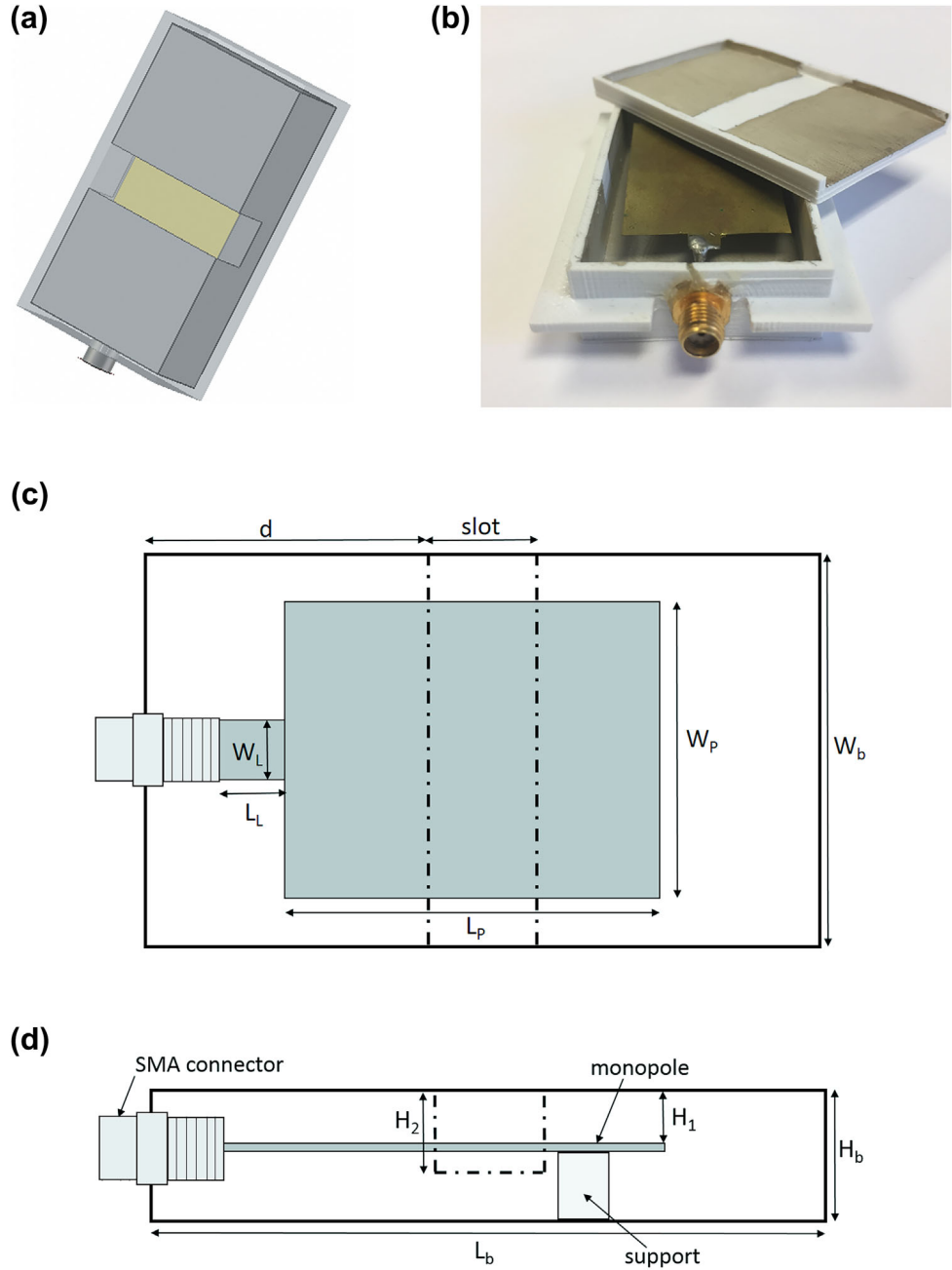

Fig. 3. (a) 3D Model of slot antenna, (b) Final antenna design, (c) Top view, and (d) Side view.

Electrodag 479SS, Henkel, Sydney, Australia) ($\sigma = 4.3 \times 10^6$ S/m). The resonant slot was not coated. It was designed following the procedure described in Varnoosfaderani et al. [2015a], where the effects of the slot dimensions were reported. The internal box dimensions were $56 \times 33 \times 11$ mm, and the wall thickness was 1.5 mm. A rectangular monopole made of brass (thickness 0.1 mm and $\sigma = 1.59 \times 10^7$ S/m) was used to excite the slot. A PLA support (PLA filament 1.75 mm, Bq, Navarra, Spain) ($5 \times 5 \times 6.7$ mm) was used to position the monopole at a fixed height. An SMA connector (SMA 8500–0000, RF Shop, Lonsdale, Australia) was soldered to the monopole and attached to the box. The 3D model of the antenna is shown in Figure 3a and the final antenna design in Figure 3b. Top and side views are shown in Figures 3c and d, respectively. The length of the slot was λ/2 (λ is the wavelength of radiation on the surface of the medium) and the monopole $\lambda_0/4$ ($\lambda_0$ is the free space wavelength), which were determined by

$$\lambda = \frac{c}{f\sqrt{\varepsilon_{eff}}} \qquad \varepsilon_{eff} \cong \frac{\varepsilon_r + 1}{2} \qquad (5)$$

where $f$ is frequency, $c$ is speed of light, $\varepsilon_{eff}$ is effective permittivity, and $\varepsilon_r$ is relative permittivity. The length of the λ/2 slot on the surface of the PLA material was initially calculated as 38.5 mm, and after optimization, the length was 47 mm long and 9 mm wide. As the length of the slot was bigger than the width of the box, it was folded onto the side walls perpendicular to the major axis (see Fig. 3).

**TABLE 1. Optimized Dimensions of the Slot Antenna Shown in Figure 3**

| Parameter | $W_b$ | $L_b$ | $W_p$ | $L_p$ | $W_L$ | $L_L$ | $d$ | $H_1$ | $H_2$ | $H_b$ |
|---|---|---|---|---|---|---|---|---|---|---|
| Value (mm) | 33 | 56 | 27 | 34 | 5 | 5.5 | 23 | 4 | 7 | 11 |

Commercial electromagnetics software with a human body model [CST Microwave Studio, 2016] was used to optimize the slot, to simulate antenna performance in free space, and when placed on a human body. The program provided the power absorbed in biological tissues. The optimization of the antenna was performed in order to improve radiation efficiency in the frequency range of interest. Table 1 shows the parameters of the antenna and box after optimization.

The voxel body model included in the CST software was truncated to reduce computational time. The model represents a 38-year-old male with height of 176 cm and weight of 69 kg. Biological material properties were recalculated using the 4-Cole-Cole formulation at the specified frequency [IFAC, 1997]. The bottom of the box was placed on different parts of the arm, torso, and thigh with no air gap between the body and antenna. Figure 4 shows an example of the box placed on the arm above the elbow of the human model.

## RESULTS

### Radiation Absorption

Figure 5 shows the relationship between antenna radiation efficiency and power absorbed in the body tissues. The antenna was placed on 10 locations of the body to achieve different values of efficiency. The input power was 100 mW at 2.45 GHz. The calculated power deposited in human tissues was found to be linearly related to the antenna efficiency (Pearson's correlation coefficient $r = 0.99$).

The radiation efficiency was 97% in free space, and decreased when the antenna was on the body with values between 62% (on the upper arm) and 75% (on the side of the torso). On other parts of the arm, the efficiency was between 67% and 72%, on the thigh it was 64%, and on the middle of the torso 68%. The maximum power absorbed in the body was 33.9 mW corresponding to the lowest performance of the antenna.

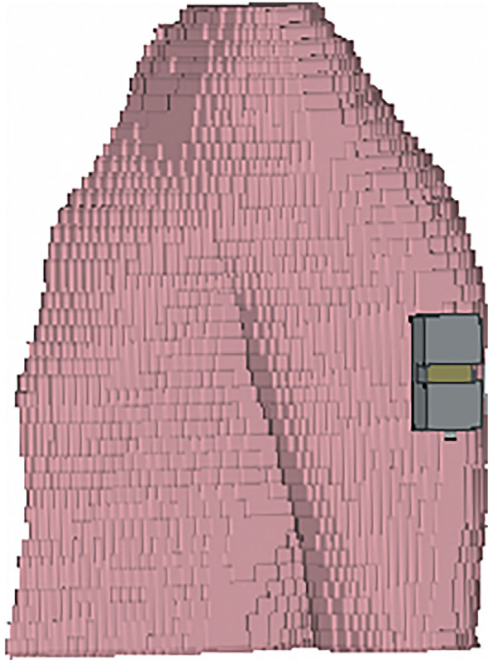

Fig. 4. Antenna placed on the arm above the elbow of the human model.

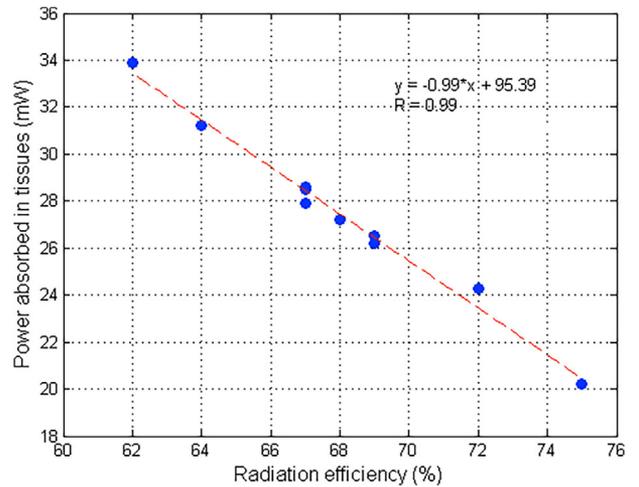

Fig. 5. Power absorbed in tissues as function of antenna radiation efficiency.

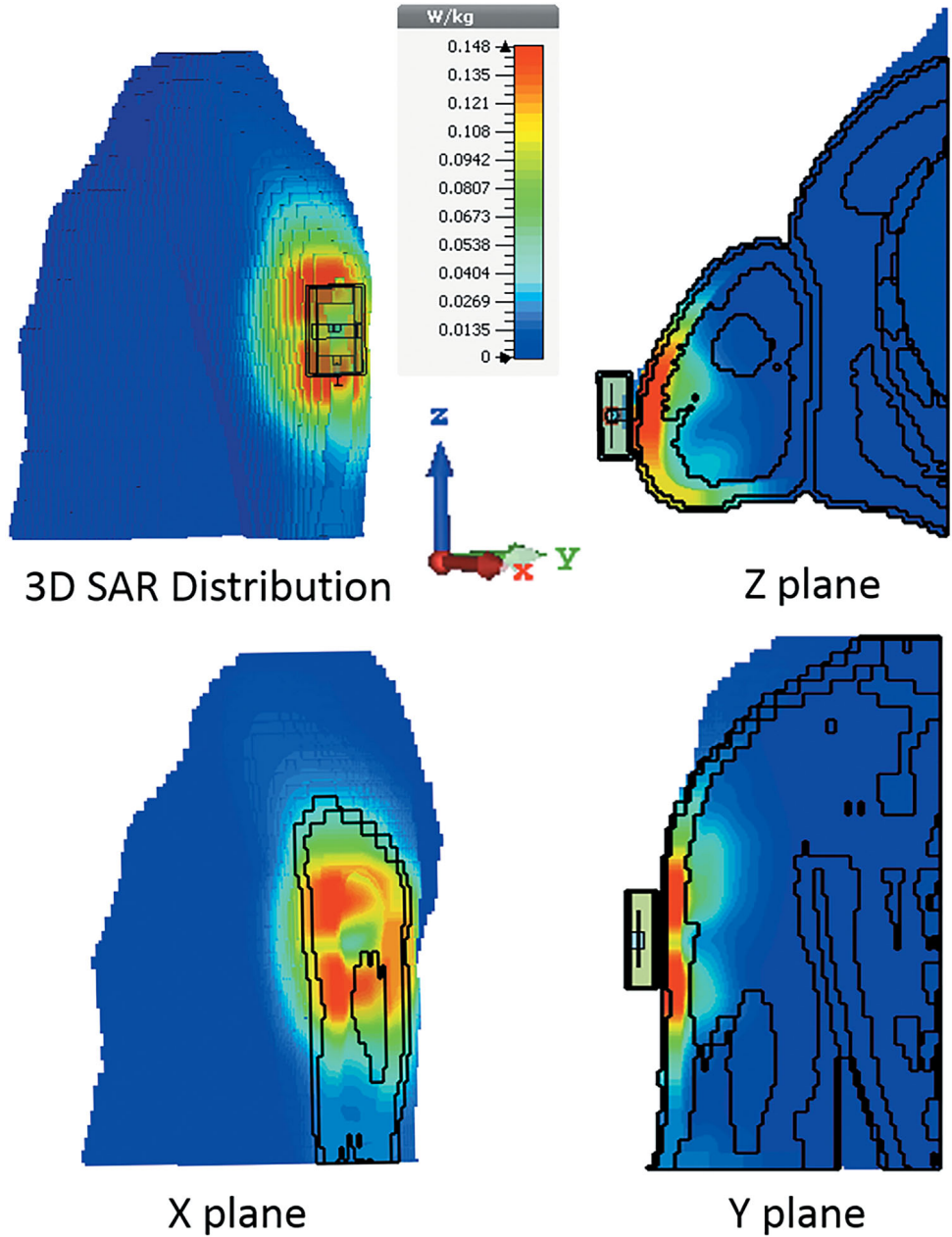

Fig. 6. SAR distribution averaged over 10 g when the antenna was placed on the arm above the elbow.

The SAR at 2.45 GHz was averaged over 10 g of mass. The maximum value was 0.316 W/kg when the antenna was located on the upper arm. The maximum SAR for other parts of the body were 0.165 W/kg on the wrist, 0.148 W/kg above the elbow, 0.196 W/kg on the thigh, and 0.185 and 0.147 W/kg on the middle and side areas of the torso. All values are well below the basic restrictions provided by international standards; at 2.45 GHz the basic restrictions for the general public are 2 W/kg for the head and trunk and 4 W/kg for the limbs when SAR is averaged over 10 g of tissue, as indicated by IEEE [2005] and ICNIRP [1998] guidelines. Figure 6 shows an example of SAR distribution in 3D and in three orthogonal cuts. The cuts are made through the maximum 10 g SAR point.

To evaluate the behavior of the antenna on different parts of the body, a figure of merit F proposed in Anguera et al. [2012] was used. This parameter defines the ratio of antenna efficiency over the SAR for a given frequency. The antenna efficiency considers the radiation efficiency and mismatch losses. The best performance (highest figure of merit) was found for locations on the left side of the torso, as evident in Figure 7. This means that in this location, the power radiated out from the body over the SAR is maximized. The smallest F value occurred when the antenna was placed on the upper arm and on the thigh, since these positions resulted in the highest SAR values. When the antenna was placed on these two parts, the results showed the highest power absorbed by the body, and therefore the lowest radiation efficiencies.

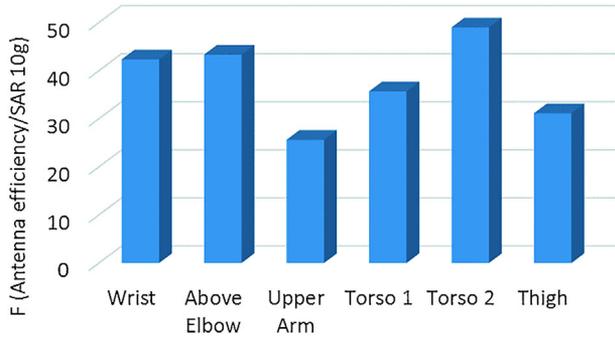

Fig. 7. Figure of merit F (Antenna efficiency over 10 g averaged SAR) at 2.45 GHz calculated at different locations of the body.

TABLE 2. Mean and Standard Deviation of Measurements, Simulation Results and Their Probability

| Position | Frequency (GHz) | | | |
| --- | --- | --- | --- | --- |
| | Mean | STD | Modeled | Probability |
| Free space | 2.390 | | 2.422 | |
| Wrist | 2.386 | $2.46 \times 10^{-2}$ | 2.282 | $2.11 \times 10^{-3}$ |
| Above elbow | 2.413 | $2.45 \times 10^{-2}$ | 2.262 | $9.43 \times 10^{-8}$ |
| Upper arm | 2.411 | $3.38 \times 10^{-2}$ | 2.307 | $1.04 \times 10^{-1}$ |
| Torso 1 | 2.438 | $6.67 \times 10^{-3}$ | 2.320 | $1.01 \times 10^{-66}$ |
| Torso 2 | 2.384 | $2.60 \times 10^{-2}$ | 2.302 | $1.02 \times 10^{-1}$ |
| Thigh | 2.425 | $1.85 \times 10^{-2}$ | 2.280 | $1.04 \times 10^{-12}$ |

**Frequency Shift**

Experimental measurements were performed on six participants to study the frequency shift when the antenna was in free space and placed on different parts of the body. The six participants had BMIs between 17 and 29 kg/m² and ages between 22 and 59 years old. The antenna was placed on various parts of the body (Fig. 2) corresponding to the simulations. The antenna was attached to the participants using plastic film. A portable Vector Network Analyzer (N9923A Field Fox Handheld RF VNA @ 6 Hz, Keysight Technologies, Santa Rosa, CA) with 50 Ω impedance was used to measure the frequency shift in each situation. The −10 dB bandwidth of the antenna in free space was 12.5% in simulations (2.27–2.57 GHz) and measurements, and did not change when the antenna was on the body. The resonant frequency changed between 2.26 and 2.32 GHz in simulations on the body and between 2.36 and 2.45 GHz in measurements. An example of the simulated and measured return loss when the antenna was in free space and on the human body is shown in Figure 8. In this case, the antenna was above the elbow and the participant had a BMI of 23.57 kg/m². The measured resonant frequency in free space was less than the frequency calculated in the simulated result. This is thought to be due to small differences between the fabricated antennas and simulation design. This includes variations in material properties, like those due to ink thickness and curing process of the conductive silver ink. Moreover, a procedure based on measurements and simulations was followed to establish the relative permittivity of the PLA box, which resulted in $\varepsilon_r = 4$ for simulations. Several monopoles of different lengths were used to feed the PLA box with a slot, and comparisons between measurements and simulations with different permittivities were performed. Small differences in box dimensions could occur, as the 3D printer had a tolerance close to 0.5 mm. When placing the antenna on the body, some deviations were observed due to different body properties (anatomical, different dielectric properties). Finally, deviations in the simulation procedure due to mesh setting were observed, as the design was meshed in the range $\lambda_0/15 - \lambda_0/20$ for the voxel model.

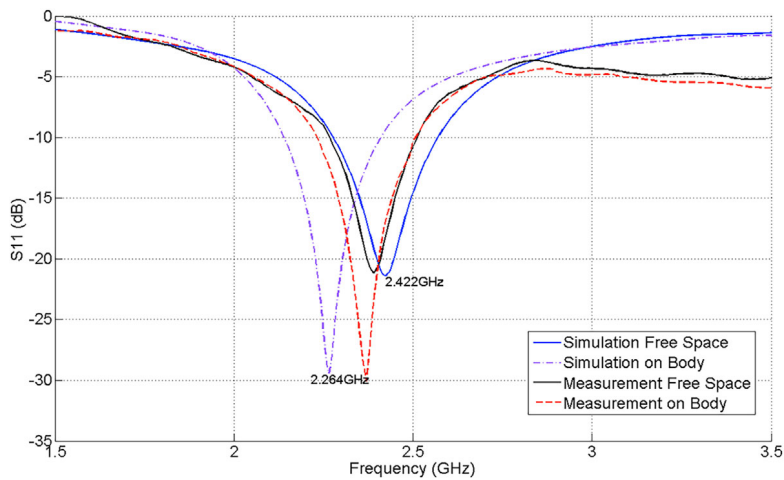

Fig. 8. Measured and simulated $S_{11}$ parameter relative to 50 Ω of antenna in free space and placed above elbow.

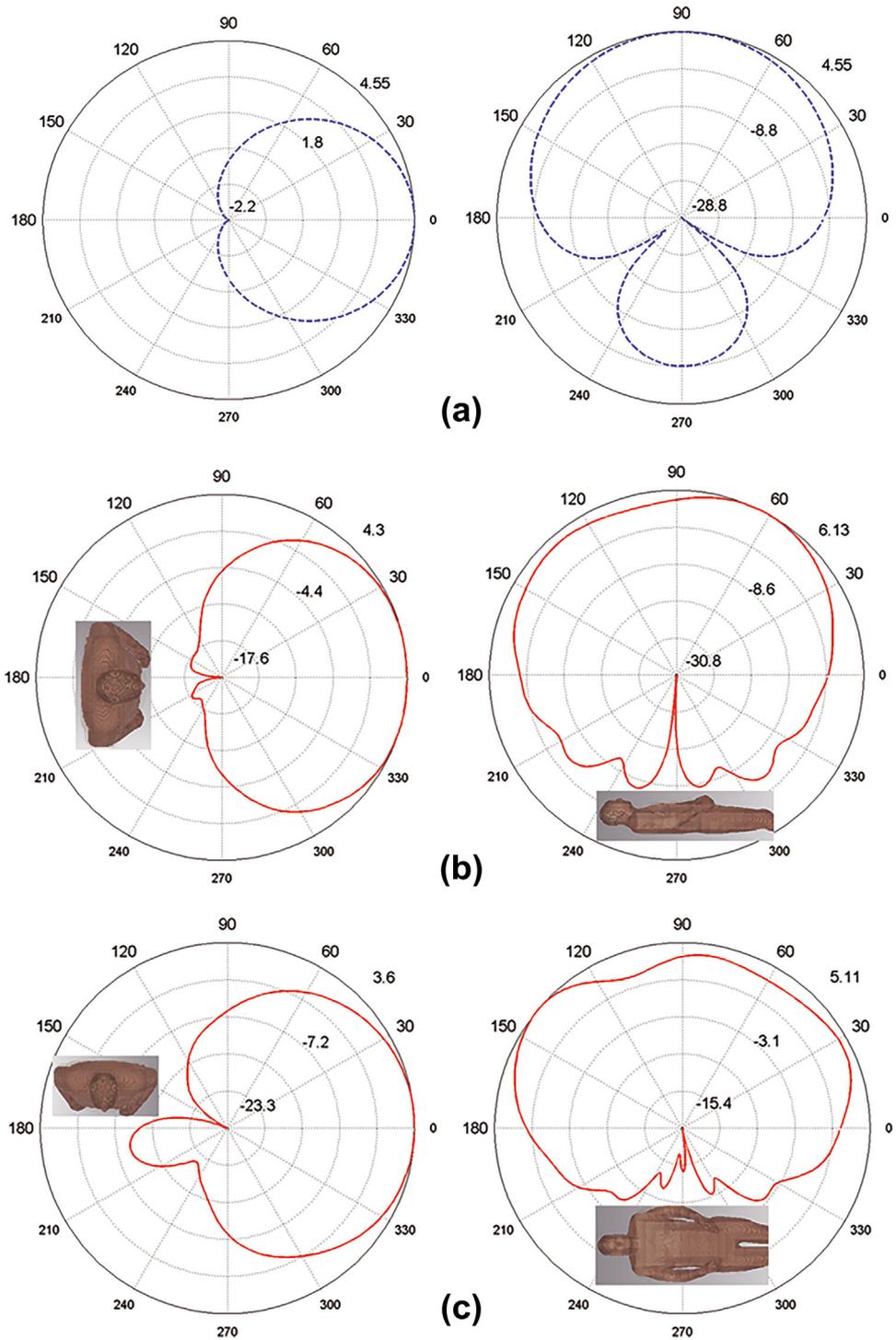

Fig. 9. Simulated radiation patterns at 2.45 GHz in both horizontal (x-y) and vertical (x-z) planes (a) in free space, (b) when antenna is on the middle of the torso, and (c) above the elbow.

The frequency shift was found to be higher in simulations compared to that observed in the experimental measurements. The maximum difference occurred when the on-body antenna was close to the elbow (6.61%) and on the thigh (5.86%). The minimum frequency shift in simulation results was obtained in the middle of the torso (4.21%), followed by upper arm (4.75%), and the side of the torso (4.95%). In measurements, the resonant frequency varied up to 2.51%.

Table 2 shows the mean and standard deviation of the resonant frequency when the antenna was placed on different parts of the body for all participants. Although six people participated in the experiment, a total of 10 measurements on each part of the body were carried out. Four people participated on two different days. In this way, uncertainties due to different positions of the antenna and skin condition could be taken into account. Simulation results are included in the column "modeled," and the probability of being statistically identical to measurements is given by the probability mass function in the column "probability." The simulation results with a better match to measurements correspond to situations in which the antenna was placed on the upper arm and on the side of the torso. The BMIs of the participants were found to not be an influential factor on the measurement results.

Radiation patterns of the antenna in free space and on-body at 2.45 GHz are shown in Figure 9. The directivity of the antenna increased when worn on the body, and the back radiation was reduced. There was no correlation between the front-to-back isolation and the power absorbed.

## DISCUSSION AND CONCLUSIONS

One of the main problems of wearable antennas is the reduction in radiation efficiency due to power absorption in human tissues. Another drawback is the resonant frequency shift. A slot antenna in a conductive box was used to minimize the interaction between the human body and antenna. In this way, not only the human effect on the antenna performance was reduced, but also the absorption loss.

The antenna design was optimized to work at 2.45 GHz, achieving a radiation efficiency of 97% in free space and between 62% and 75% when it was placed on the body. In simulations, the resonant frequency reached its maximum shift when the antenna was above the elbow (6.61%). Experimental measurements showed a maximum frequency shift of 2.51%. Moreover, when using this antenna, design results did not depend on the body-mass index for each individual.

One limitation of this wearable antenna is that it has to be fixed and in contact with the skin to prevent changes in performance. The probe inside the box needs to be precisely positioned for maximum performance.

The specific absorption rate that was studied by means of simulations and results proved that this antenna is appropriate for on/off body communications since the maximum 10 g averaged SAR value was 0.316 W/kg for 100 mW input power. This is well below international limits and this value would be reduced if the distance from the body was increased [Karmokar et al., 2010]. These SAR values are also satisfactory in comparison to results at the same frequency reported by other authors [De Santis et al., 2012; Soh et al., 2015].

SAR values are useful to verify compliance with health standards and are representative of localized absorption. This does not allow the evaluation of power absorbed by different parts of the body, since SAR results give information about maximum absorption averaged over 10 g. The power absorbed in tissues (Eq. 2) was found to be the best parameter for measuring total absorption in parts of the body [Risco et al., 2012]. We demonstrated that two similar values of SAR can be related to very different values of head absorption. In our study, results showed that when the slot antenna was above the elbow, absorption was less correlated with SAR than at other locations.

Considering all the parameters studied in this work (radiation efficiency, frequency shift, power absorbed, and SAR), it can be concluded that this antenna performs efficiently at most locations of the body and by different people.